\documentclass[showpacs,twocolumn,preprintnumbers,showkeys,superscriptaddress,amsmath,amssymb,nofootinbib]{revtex4}
\usepackage{color}
\usepackage{amsmath}
\usepackage{amssymb}
\usepackage[utf8]{inputenc}
\usepackage{amsfonts}
\usepackage{bm}
\usepackage{bbold}

\newcommand{\be}{\begin{equation}}
\newcommand{\ee}{\end{equation}}
\newcommand{\ba}{\begin{eqnarray}}
\newcommand{\ea}{\end{eqnarray}}
\newcommand{\p}{\partial}
\def\ni{\noindent}

\begin{document}


\title{\Large A general relativity formulation for the Doplicher-Fredenhagen-Roberts noncommutative space-time}


\author{M. J. Neves} \email{mneves@ua.edu}
\affiliation{Department of Physics and Astronomy, University of Alabama, Tuscaloosa, Alabama 35487, USA}
\affiliation{Departamento de F\'{i}sica, Universidade Federal Rural do Rio de Janeiro, BR 465-07, 23890-971, Serop\'edica, RJ, Brazil}

\author{Everton M. C. Abreu}\email{evertonabreu@ufrrj.br}
\affiliation{Departamento de F\'{i}sica, Universidade Federal Rural do Rio de Janeiro, BR 465-07, 23890-971, Serop\'edica, RJ, Brazil}
\affiliation{Departamento de F\'{i}sica, Universidade Federal de Juiz de Fora, 36036-330, Juiz de Fora, MG, Brazil}
\affiliation{Programa de P\'os-Gradua\c{c}\~ao Interdisciplinar em F\'isica Aplicada, Instituto de F\'{i}sica, Universidade Federal do Rio de Janeiro, 21941-972, Rio de Janeiro, RJ, Brazil}


\date{\today}

\begin{abstract}
\ni The Doplicher, Fredenhagen and Roberts (DFR) noncommutative (NC) formalism was proposed in a curved space-time.
In DFR approach, the NC parameters are promoted to a set of space-time coordinates. Consequently,
the field theory defined on this space contains extra dimensions. We propose a metric containing the new coordinates
to explain the length measurements of this extended space. We promote the Minkowski part of this extended metric to
a metric defined in a curved manifold. Thus, we can build up a NC gravitation model with extra-dimensions. Based on the
curved metric, we propose an Einstein-Hilbert action to describe the NC general relativity model on the DFR space.  It is a new result in the literature since there is not a general relativity formalism in DFR approach.
We also introduce the NC version of the general relativistic field equations. As an application of the formalism, the weak field approximation is considered in the field equations to generate the wave equations for the graviton propagation in DFR space.
\end{abstract}



\maketitle



\section{Introduction}
The need to control both the ultraviolet and infrared (UV/IR) divergences that eventually emerge in several calculations in QFT, have motivated theoretical physicists to suggest the modification of the structure of the standard space-time. One of the possible solutions was to promote the continuum space-time of QFT to a discrete, and consequently compact, space-time with a noncommutative (NC) algebra of coordinates. In this formalism, the space-time coordinates $x^{\mu}=\left( \, x^{0}=ct \, , \, x^{1} \, , \, x^{2} \, , \, x^{3} \, \right)$ are promoted to operators $\hat{X}^{\mu}=\left( \, \hat{X}^{0} \, , \, \hat{X}^{1} \, , \, \hat{X}^{2} \, , \, \hat{X}^{3} \, \right)$. They satisfy non-trivial commutation relations in which the result is an antisymmetric constant matrix, called $\theta^{\mu\nu}$, that play a fundamental r\^ole in the theory and its dimension dwells at Planck scale, which classifies it as a semi-classical quantity, at least. Thereby, we have a kind of fuzzy space-time where the position coordinates uncertainty can introduce a fundamental length scale in the theory.
The space-time defined with these coordinates is also called NC space-time. The paper by H. S. Snyder was the first published work that considered the space-time as a NC one \cite{snyder47}.   Shortly after, C. N. Yang \cite{yang47} corrected and enlarged the invariance group of Snyder's model.   After that, this NC formalism was not discussed for years.

After the interesting result that the algebra obtained from string theory embedded in a magnetic field background brought a NC algebra, the noncommutativity (NCY) concept of space-time was rekindle \cite{seibergwitten99}. One of the ways, the most popular at least, of introducing NCY is through the Moyal-Weyl (MW) product where the NC parameter, $\theta^{\mu\nu}$, is an antisymmetric constant matrix, similar to Snyder's formalism parameter.
However, for calculations at higher orders in the perturbative series, the MW product turns out to be
highly nonlocal, and it compels us to work with low orders in $\theta^{\mu\nu}$ parameter. Besides, since the $\theta$-parameter is a Planck scale quantity, we can eliminate its higher-orders without loss of a definite result. It can also be demonstrated that the time-space coordinates ($\theta^{0i}$, space-like NCY) can cause unitarity problems \cite{unitarity,chaichianPLB}.
Although it keeps the translational invariance, the Lorentz symmetry is broken \cite{Szabo03}. As an example, for the hydrogen atom problem, since a constant parameter in the algebra means a fixed axis, it breaks the rotational symmetry of the model, and removes the degeneracy of the energy levels \cite{Chaichian}. One way to recover the Lorentz symmetry is to use the so-called Doplicher, Fredenhagen and
Roberts (DFR) formalism that introduced another kind of NC space-time. In this formalism, the NC parameter is promoted to ordinary coordinate of space-time, such that, $\theta^{\mu\nu} \mapsto \hat{\Theta}^{\mu\nu}$, where $\hat{\Theta}^{\mu\nu}$ keeps both the square length dimension and Planck scale features
\cite{DFR1,DFR2}. As a consequence, the DFR space-time is extended to ten dimensions: four relative to the Minkowski space-time and six relative to the eigenvalues of $\hat{\Theta}^{\mu\nu}$ coordinates, namely, it has the same number of dimensions as SUGRA. Posteriorly, many authors extended the DFR formalism to include a canonical momentum associated with $\theta^{\mu\nu}$ \cite{Morita,Morita2003,alexei,jhep,saxell,amorim2008,amorim2009,amorim20092,Amorim1,amo}. Besides the Lorentz invariance, the causality properties in this space must be preserved \cite{AbNeves,AbNeves2013,AbNeves2014}. Although the Lorentz symmetry is
recovered in DFR, the field theories defined on this space are unitary if we impose $\hat{\Theta}^{0i}=0$, as we said before, for the commutators between time and spatial coordinates \cite{BassetoJHEP2001,ChuNPB2002}. To sum up, we guarantee the unitarity of the model but the Lorentz symmetry is broken.  Thus, from now on, we will be dealing with a model that preserves just the translation symmetry. Hence, we have a NCY associated with the spatial coordinates, the DFR space is reduced for $4+3$ dimensions.

One of the challenges of NC formalism and its structures is to construct a NC gravity \cite{RivellesNP,Kober,Gupta,KoberCQG,Miao,Nicolini,Zarro,ChaichianPRD,Fucci,Banerjee,Marculescu,Das,Mitra}, which means that a Planck scale parameter will be part of a gravitation structure which suggest a semi-classical, at least, quantum-type formulation of gravity.
In this letter, we will propose a gravitation model defined on DFR space of $4+3$ dimensions \cite{Amorim1,Amorim2,Amorim3,Amorim4,Amorim5}. This gravitation model is motivated by many extensions of gravity theories in higher dimensions, like
SUGRA \cite{DeWit,Nath,Brandt2002,Martin2016}. The introduction of NC effects in gravity is also a subject that has a massive literature \cite{reviews,szabo2006,szabo2009,grav,Calmet,Rivelles2006,Rivelles2003,Steinacker2007,Steinacker2009,Banerjee2005,Douglas1998,Horvathy2002,Horvathy2004,Horvathy2005}.  The main motivation of DFR formalism is to analyze gravity effects in a NC space-time. However, a formal structure, with the general relativity main ingredients was not provided so far.

In this paper we have proposed a Einstein-Hilbert action in $4+3$ dimensions and the corresponding gravitational field equations in DFR framework. As a particular case, the NC gravity in four dimensions must be contained in the equations for this extended space-time. The commutative limit is recovered when
the length scale of the theory goes to zero, and the usual Einstein-Hilbert action and the field equations from the general relativity are reobtained.  As an application of the formalism, by using the linear approximation of the gravitational field, we obtain the wave equation for the graviton on the DFR space-time.

We have considered the following organization of the ideas here. In section \ref{sec2}, we provided a very brief review of DFR main points. In section \ref{sec3}, we introduce the main ingredients of DFR general relativity structure. In section \ref{sec4}, the conclusions and perspectives were described.


%
\section{The NC DFR framework}
\label{sec2}

The original DFR formalism has the coordinates of the NC antisymmetric
parameter $\theta^{\mu\nu}=\left( \, \theta^{01},\theta^{02},\theta^{03},\theta^{12},\theta^{13},\theta^{23} \, \right)$ acting as coordinates of this ten dimensional space-time. The other four dimensions come from the Minkowski space-time.  However, in terms of a quantum formulation, all these coordinates are promoted to operators, so that we have,
$\hat{\Theta}^{\mu\nu}=\left(\hat{\Theta}^{01},\hat{\Theta}^{02},\hat{\Theta}^{03},\hat{\Theta}^{12},\hat{\Theta}^{13},\hat{\Theta}^{23}\right)$.
In this letter, as we said before, to avoid unitarity problems, we will consider that the time-space coordinates are zero, {\it i.e.}, $\hat{\Theta}^{0i}=0$, which means that our NC approach is purely spatial.
This statement solves any unitarity problems that could emerge from the corresponding effective gravitation theory, which is no more dimensionally analogous to SUGRA. Thereby, we have a NC space-time of $D=4+3$ dimensions. The extended observable DFR algebra is represented by the following commutators
\begin{eqnarray}\label{algebraDFR}
\left[ \, \hat{X}^{i} \, , \, \hat{X}^{j} \, \right] \!\!&=&\!\! i \, \varepsilon^{ \, ijk} \, \hat{\Theta}^{k}
\; , \qquad
\left[ \, \hat{X}^{i} \, , \, \hat{\Theta}^{j} \, \right] = 0
\; , \qquad \qquad \quad
\nonumber \\
\left[ \, \hat{\Theta}^{i} \, , \, \hat{\Theta}^{j} \, \right] \!\!&=&\!\! 0
\; , \qquad \qquad \quad
\left[ \, \hat{X}^{\mu} \, , \, \hat{P}^{\nu} \, \right] = i\,\eta^{\mu\nu} \, \hat{\mathbb{1}}
\hspace{0.2cm} , \hspace{0.2cm}
\nonumber \\
\left[ \, \hat{P}^{\mu} \, , \, \hat{P}^{\nu} \, \right] \!\!&=&\!\! 0
\hspace{0.2cm} , \hspace{0.2cm} \qquad \qquad
\left[ \, \hat{\Theta}^{i} \, , \, \hat{P}^{\rho} \, \right] = 0
\hspace{0.2cm} , \hspace{0.2cm}
\nonumber \\
\left[ \, \hat{P}^{\mu} \, , \, \hat{K}^{i} \, \right] \!\!&=&\!\! 0
\; , \qquad \qquad \quad
\left[ \, \hat{\Theta}^{i} \, , \, \hat{K}^{j} \, \right] = i \, \delta^{ij} \, \hat{\mathbb{1}} \; ,
\nonumber \\
\left[\, \hat{X}^{i} \, , \, \hat{K}^{j} \, \right] \!&=&\! \frac{i}{4} \, \varepsilon^{ijk} \, \hat{P}^{k} \; , \;\quad
\left[ \, \hat{X}^{0} \, , \, \hat{\Theta}^{i} \, \right] = 0 \;,
\end{eqnarray}
where we have adopted the Minkowski metric $\eta^{\mu\nu}=\mbox{diag}(+---)$.
By convenience of notation, we have used the dual operators of $\hat{\Theta}^{ij}$ and
$\hat{K}^{ij}$ as our NC operators, namely,
$\hat{\Theta}^{ij}=\varepsilon^{ \, ijk} \, \hat{\Theta}^{k}$ and $\hat{K}^{ij}=\varepsilon^{\, ijk} \, \hat{K}^{k}$, respectively.
The $\hat{K}^{i}$ operator is the conjugated canonical momentum associated with the coordinate operator
$\hat{\Theta}^{i}$. Here we have adopted natural units $c=\hbar=1$, where the $\theta^{i}$-coordinates has squared length dimension,
and the $k^{i}$ momentum has inverse of the squared length dimension.
The components of rotation and boost generators
$\hat{\Sigma}^{\mu\nu}=\left( \, \hat{\Sigma}^{0i} \, , \, \hat{\Sigma}^{ij}  \, \right)$ in the
Lorentz group are defined by
\begin{eqnarray}\label{Mmunu}
\hat{\Sigma}^{0i} &=& \hat{X}^{0} \, \hat{P}^{i}
- \hat{\xi}^{i} \, \hat{P}^{0} \; ,
\nonumber \\
\hat{\Sigma}^{ij} &=& \hat{\xi}^{i} \, \hat{P}^{j}
- \hat{\xi}^{j} \, \hat{P}^{i}
- \hat{\Theta}^{i} \, \hat{K}^{j}
+ \hat{\Theta}^{j} \, \hat{K}^{i} \; ,
\end{eqnarray}

\ni where $\hat{\xi}^{i} = \hat{X}^{i} \, + \, i \, \varepsilon^{ijk} \, \hat{\Theta}^{j} \, \hat{P}^{k}/2$
is the shifted operator $\hat{X}^{i}$, which characterizes the well known Bopp shift \cite{Gamboa,Kokado,Kijanka,Calmet1,Calmet2}.
%

%
%
%
%
%
%
%
%
With the help of Eq. (\ref{algebraDFR}), we can construct the commutation relations
%
\begin{eqnarray}\label{algebraPoincareDFR}
\left[ \, \hat{\Sigma}^{\mu\nu} \, , \, \hat{P}^{\rho} \right] &=& \, i \, \big( \, \eta^{\mu\rho} \, \hat{P}^{\nu}
-\eta^{\nu\rho} \, \hat{P}^{\mu} \, \big)
\; , \nonumber \\
\left[ \, \hat{\Sigma}^{0i} \, , \, \hat{K}^{j} \right] &=& 0
\; , \nonumber \\
\left[ \, \hat{\Sigma}^{ij} \, , \, \hat{K}^{k} \right] &=& \, i \, \big( \, \delta^{jk} \, \hat{K}^{i}
-\delta^{ik} \, \hat{K}^{j} \, \big)
\; ,
\nonumber \\
\left[ \, \hat{\Sigma}^{\mu\nu} \, , \, \hat{\Sigma}^{\rho\sigma} \, \right] &=& i\Big( \, \eta^{\mu\sigma} \, \hat{\Sigma}^{\rho\nu}
-\eta^{\nu\sigma} \, \hat{\Sigma}^{\rho\mu}-\eta^{\mu\rho} \, \hat{\Sigma}^{\sigma\nu} \nonumber \\
&&\quad +\,\,\eta^{\nu\rho} \, \hat{\Sigma}^{\sigma\mu} \, \Big) \; .
\end{eqnarray}
%
Consequently, the operators $\hat{P}^{\mu}$, $\hat{\Sigma}^{\mu\nu}$ and $\hat{K}^{i}$ close the DFR Poincar\'e algebra
of generators of translations, rotations and Lorentz boosts. The element of the algebra in Eq. (\ref{algebraPoincareDFR}),
that commutes with all the generators $\left( \, \hat{P}^{\mu} \, , \, \hat{\Sigma}^{\mu\nu} \, , \, \hat{K}^{i} \, \right)$
is defined by the first Casimir operator. Using the Schur's Lemma, the first Casimir operator is proportional
to the
squared mass $m^2$ of the particle: $\hat{P}_{\mu} \, \hat{P}^{\mu} \, - \, \lambda^{2} \, \hat{K}^{i} \, \hat{K}^{i}=m^2 \,
\hat{\mathbb{1}}$.
%
%
%
%
Using the coordinate representation, the operators $\hat{P}^{\mu}$ and $\hat{K}^{i}$
can be written in terms of the derivatives
%
$\hat{P}_{\mu} \longmapsto i \, \p_{\mu}$
and
$\hat{K}_{i} \longmapsto i \, \partial_{{\bm \theta}}$.
%
Thereby, the first Casimir operator on-shell result leads us to the field equations
of scalar and fermions fields \cite{EMCAbreuMJNeves20171}. From now on,
the notation ${\bm \theta}$ means the $\theta^{i}$-coordinates
${\bm \theta}=\left( \, \theta^{1} \, , \, \theta^{2} \, , \, \theta^{3} \, \right)$ and
$\partial_{{\bm \theta}}=( \, \p_{\theta_{1}} \, , \, \p_{\theta_{2}} \, , \, \p_{\theta_{3}} \, )$ is the derivative operator
relative to the coordinates $\left( \, \theta^{1} \, , \, \theta^{2} \, , \, \theta^{3} \, \right)$.
The length scale $(\lambda)$ was introduced in the dispersion relation to keep the squared mass dimension of the operator
$\hat{K}^{i}\hat{K}^{i}$. However, there is a reason to introduce this parameter connected to the
extra dimensions of the coordinate $\hat{\Theta}^{i}$ which we will see in a moment.


Concerning the diffeomorphism invariance, it is one of the most underlying issues of any gravity theory.
We had recent advances in NC gravity \cite{szabo}, but there is still no unique and totally
adequate path to understand the full diffeomorphism group on a NC manifold.
As an alternative, we can apply the Seiberg-Witten approach \cite{seibergwitten99} where all symmetries are reduced, including the
diffeomorphisms, to the commutative ones.  But we have to redefine all the nonlinear fields.
Nevertheless, the computations beyond the noncommutativity parameter leading order  are
very difficult to achieve in this formalism \cite{Chamseddine,Garcia,ChaichianPLB2008}. Another way to extend the
diffeomorphism transformations to NC spaces is to make their action twisted \cite{Aschieri05,Aschieri06}. In this way, we can
design a whole twisted-invariant NC gravity action having just the correct number
of symmetries. On the other hand, the twisted symmetries are not considered as real physical symmetries.
Hence, we are not allowed to use them to be free of any degrees of freedom.  In our case, we are dealing
with a seven dimensional theory where the coordinates have area dimension, these features are even hard
to deal.  It is an ongoing research that will be published elsewhere.


Since we have all the symmetries described by the algebra in Eq. (\ref{algebraPoincareDFR})
and the dispersion relation of the particles are well defined, we can construct the DFR action.
The action to describe an arbitrary field $\phi$ defined in the DFR
space is given by the action
\begin{eqnarray}\label{actionS}
{\cal S}(\phi)=\int d^{4}x \, d^{3}{\bm \theta} \, W({\bm \theta}) \,
{\cal L}(\phi\star,\partial_{\mu}\phi\star,\partial_{{\bm \theta}}\phi\star) \; ,
\end{eqnarray}
where the volume element contains the ${\bm \theta}$-integration measure $W$ as
a function of the three  ${\bm \theta}$-coordinates. The $W$ function was introduced in the integration
of NC field theory to smooth the divergences of the integration in the extra ${\bm \theta}$-space.
It has two basic properties: (i) it should be an even function of the  ${\bm \theta}$-coordinates,
namely, $W(-\,{\bm \theta})\,=\,W({\bm \theta})$, which implies that the integration in the ${\bm \theta}$-space
is isotropic; and (ii) for large coordinates ${\bm \theta}$, it zeroes quickly so that all ${\bm \theta}$-integrals
are well defined. Hence, a normalization condition is assumed when integrated in the ${\bm \theta}$-space
\cite{Carlson,Morita,Morita2003,Conroy2003,Saxell}. Following these properties, the simplest $W$ function has the Gaussian form
\begin{eqnarray}
W({\bm \theta})=\left(\frac{1}{2\pi\lambda^{2}} \right)^{3} \, \mbox{exp}\left(-\frac{ {\bm \theta}^{2} }{4 \, \lambda^{4}}\right) \; .
\end{eqnarray}
The length parameter $\lambda$ can be interpreted as the NC scale energy $(\Lambda_{NC})$
in terms of the expected value of the operator $\hat{\Theta}^{2}=\hat{\Theta}^{i}\hat{\Theta}^{i}$:
%
$\Lambda_{NC}=(12/\langle \hat{\Theta}^{2} \rangle)^{1/4} \equiv \lambda^{-1}$.
%
In NC quantum electrodynamics, some processes yield the range of $ \Lambda_{NC} \gtrsim 0.1 - 1.7$ TeV.
The Bhabha scattering, dilepton and diphoton production in the LEP data have the lower bound of
$\Lambda_{NC} \gtrsim 160$ GeV with 95 \% C.L. \cite{Carone,Conroy2003}. The light-by-light scattering yields
a lower bound of $\Lambda_{NC} \gtrsim 0.1$ TeV \cite{horvat2020}. The theoretical limit from Lamb shift
in the hydrogen atom is $\Lambda_{NC} \gtrsim 10$ TeV \cite{AkoforPRD2009}.
An important point in the DFR scenario is that the product between two fields as function
of NC variables keeps the usual form of the MW product. The Weyl symbol provides
a map from the operator algebra to the functions algebra equipped with a star-product $\star$
via the MW correspondence
%
%
in which the star-product $\star$ is defined by
\begin{eqnarray}\label{ProductMoyal}
\left. f(x,{\bm \theta}) \star g(x,{\bm \theta}) =
e^{\frac{i}{2} \, {\bm \theta} \cdot (\nabla \times \nabla^{\prime})}
f(x,{\bm \theta}) \, \, g(x^{\prime},{\bm \theta}) \right|_{x^{\prime}=x} \, , \;
\end{eqnarray}
for any arbitrary functions $f$ and $g$ of the coordinates $(x^{\mu},{\bm \theta})$, with $\nabla$ and $\nabla^{\prime}$
being gradient operators in relation to $x$ and $x^{\prime}$, respectively. In both sides of Eq. (\ref{ProductMoyal}) we have that $f$ and $g$
are NC functions since they depend on the coordinate $x^{\mu}$. If $f$ depends on $\left( \, x \, , \, {\bm \theta} \, \right)$,
and $g$ depends only on ${\bm \theta}$, the Moyal product between these functions is reduced to the usual product, {\it i.e.},
$f(x,{\bm \theta})\star g({\bm \theta})=f(x,{\bm \theta}) \, g({\bm \theta})$, because ${\bm \theta}$ commutes with $x^{\mu}$,
and with itself.
The integration measure in Eq. (\ref{actionS}) suggests us to construct a general metric that encompasses the Minkowski
space-time and the extra space associated with the ${\bm \theta}$-coordinates. Thereby, we propose the line element
of this extended space-time into the form
\begin{eqnarray}\label{dsDFR}
ds^{2}=\Xi_{AB}({\bm \theta}) \, dX^{A} \star dX^{B} \; ,
\end{eqnarray}
where the coordinates $X^{A}$ are redefined by $X^{A}=\left\{ \, x^{\mu} \, , \, {\bm \theta}/(2\pi\lambda^2) \, \right\}$,
where $\mu = \left\{ \, 0 \, , \, 1 \, , \, 2 \, , \, 3 \, \right\}$ and $A=\left\{ \, 0 \, , \, \ldots \, , \, 7 \, \right\}$.
Note that the coordinates of ${\bm \theta}$ in $X^{A}$ are defined as being dimensionless due to the length scale $\lambda$.
The components of the metric $\Xi_{AB}({\bm \theta})$ are elements of the $7 \times 7$ diagonal matrix \cite{EMCAbreuMJNeves20171}
\begin{equation}\label{metricG}
\Xi_{AB}({\bm \theta})=\mbox{diag}\left( \, \eta_{\mu\nu} \, , \, e^{-\frac{\theta_{1}^{\, 2}}{2\lambda^{4}}}
\, , \, e^{-\frac{\theta_{2}^{\, 2}}{2\lambda^{4}}}
\, , \, e^{-\frac{\theta_{3}^{\, 2}}{2\lambda^{4}}}
\, \right) \; .
\end{equation}
The inverse of the metric in Eq. (\ref{metricG}) is defined by $\Xi_{AB}({\bm \theta}) \, \Xi^{BC}({\bm \theta})=\delta_{A}^{\; \; \, C}$,
where the delta on the right side can be represented by the identity matrix $7 \times 7$.
With this definition, the $W$ function can be written in terms of the determinant of $\Xi({\bm \theta})$ matrix
$d^{4}x \, d^{3}{\bm \theta} \, W({\bm \theta})=d^{4}X \, \sqrt{-\Xi}$. The commutative limit
in  Eq. (\ref{metricG}) can be obtained when $\lambda \rightarrow 0$ through the identity
\begin{eqnarray}\label{DeltaID}
\lim_{\lambda \rightarrow 0} \, \frac{e^{-\frac{{\bm \theta}^{\,2}}{2 \, \lambda^{4}}}}{(2\pi\lambda^{2})^{3}} = \delta^{(3)}\left({\bm \theta}\right) \; ,
\end{eqnarray}
where the ${\bm \theta}$-part in Eq. (\ref{dsDFR}) is zero due to $\delta^{(3)}$-Dirac property,
{\it i.e.}, $\delta^{(3)}\left({\bm \theta}\right) \, d^{3}{\bm \theta}=0$.
Thus, it is direct to check that the line element in Eq. (\ref{dsDFR}) recovers the
usual Minkowski space-time, and the action in Eq. (\ref{actionS}) is reduced to the 4D QFT commutative case.
%


\section{The DFR gravitation framework}
\label{sec3}
Following the definition of the extended metric $\Xi_{AB}({\bm \theta})$, we generalize the line element in Eq. (\ref{dsDFR})
such that its Minkowski part can be modified for a curved space-time given by
\begin{eqnarray}\label{dsDFRgx}
ds^{2}=G_{AB}(x,{\bm \theta}) \star dX^{A} \star dX^{B} \; ,
\end{eqnarray}
where $\eta_{\mu\nu}$ is promoted to $g_{\mu\nu}(x)$ and $G_{AB}(x,{\bm \theta})$ describes the space of
a manifold attached to the extra ${\bm \theta}$ space
\begin{equation}\label{metricGx}
G_{AB}(x,{\bm \theta})=\mbox{diag}\left(g_{\mu\nu}(x) , e^{-\frac{\theta_{1}^{\, 2}}{2\lambda^{4}}} , e^{-\frac{\theta_{2}^{\, 2}}{2\lambda^{4}}} , e^{-\frac{\theta_{3}^{\, 2}}{2\lambda^{4}}} \right) \; .
\end{equation}
The inverse of this metric is extended to the Moyal-product if there exists a matrix $G^{AB}(x,{\bm \theta})$,
such that
\begin{eqnarray}
G_{AB}(x,{\bm \theta}) \star G^{BC}(x,{\bm \theta})=\delta_{A}^{\; \; \, C} \; .
\end{eqnarray}
Using the extension from the metric in Eq. (\ref{metricGx}), we propose the Einstein-Hilbert action in the DFR framework such as
\begin{eqnarray}\label{SGDFR}
S_{EH}(g_{\mu\nu})=-\frac{1}{\kappa^2} \int d^{4}X \, \sqrt{-G} \star R \; ,
\end{eqnarray}
where $\kappa= \sqrt{32\,\pi\, G_{N}} \simeq 8.15 \times 10^{-5}$m, is the coupling constant written in terms of the
gravitational constant $G_{N}\,=\,6.674 \times 10^{-11} \, \mbox{m}^2$. The Ricci scalar is defined by $R=G_{AB}(x,{\bm \theta}) \star R^{AB}$,
in which the Ricci tensor $R^{AB}$ can be written as
\begin{eqnarray}\label{Riccitensor}
R_{AB} &=& \partial_{A}\Gamma^{C}_{\ \ C B} - \partial_{C}\Gamma^{C}_{\ \ AB} \nonumber \\
&& + \,\frac{1}{2} \left( \, \Gamma^{C}_{\ \ A D} \star \Gamma^{D}_{\ \ C B} + \Gamma^{D}_{\ \ C B} \star \Gamma^{C}_{\ \ A D} \, \right) \nonumber \\
&& - \,\Gamma^{C}_{\ \ AB} \star \Gamma^{D}_{\ \ D C} \; ,
\end{eqnarray}
and $\Gamma^{C}_{\ \ AB}$ is the affine connection of this space
\begin{equation}\label{Gammag}
\Gamma^{C}_{\,\,\, AB} = \frac{1}{2} \, G^{C D} \star \left( \, \partial_{A} G_{B D} + \partial_{B} G_{A D} - \partial_{D} G_{AB} \, \right) \; .
\end{equation}
The derivative operators with capital index in Eqs. (\ref{Riccitensor}) and (\ref{Gammag}) means the derivative with relation to the $X^{A}$ coordinates (variables in DFR space) defined previously, namely, $\partial_{A}=\partial/\partial X^{A}=\left( \, \partial_{\mu} \, , \, \lambda \, \partial_{\theta_{i}}  \, \right)$. Since we have all the tensor structure in this extended space, we propose an Einstein's field equation for DFR as being
\begin{eqnarray}\label{EqEinstein}
R_{AB}-\frac{1}{2} \, G_{AB}(x,{\bm \theta}) \star R\,=\,-\,8\pi G_{N} \, T_{AB} \; ,
\end{eqnarray}
where $T_{AB}$ is the energy-momentum tensor for matter fields (scalars, spinors, vectors and etc.)
with the components $T_{AB}=\left( \, T_{\mu\nu} \, , \, T_{\mu \theta_{i}} \, , \, T_{\theta_{i}\theta_{j}} \, \right)$.
Note that $T_{\mu \theta_{i}}$ and $T_{\theta_{i}\theta_{j}}$ are the new components due to the extra $\theta$-dimension.
As an example, the energy-momentum tensor of a NC scalar field $\phi$ on the DFR space is given by
\begin{equation}\label{TAB}
T_{AB} = \frac{1}{2} \left( \, \partial_{A}\phi \star \partial_{B}\phi
+\partial_{B}\phi\star\partial_{A}\phi  \, \right)
-G_{AB}(x,{\bm \theta}) \star {\cal L}_{\phi} \; ,
\end{equation}
where ${\cal L}_{\phi}$ is the Lagrangian of a scalar field $\phi$ with mass $m$
\begin{eqnarray}
{\cal L}_{\phi}=\frac{1}{2} \, G_{AB}(x,{\bm \theta}) \star \left( \, \partial^{A}\phi \star \partial^{B}\phi \, \right)
- \frac{1}{2} \, m^{2} \, \phi \star \phi \; .
\end{eqnarray}
Note that the connection, the Ricci tensor and the energy-momentum tensor are symmetric by exchanging the indexes
$A \leftrightarrow B$. This symmetry is kept at the commutative limit.
%
%
More explicitly, the components of $T_{AB}=\left( \, T_{\mu\nu} \, , \, T_{\mu\theta_{i}} \, , \, T_{\theta_{i}\theta_{j}}  \, \right)$
can be written as
\begin{eqnarray}
T_{\mu\nu} \!&=&\! \frac{1}{2} \left( \, \partial_{\mu}\phi \star \partial_{\nu}\phi+\partial_{\nu}\phi \star \partial_{\mu}\phi \, \right)
-g_{\mu\nu}(x) \star {\cal L}_{\phi} \; ,
\nonumber \\
T_{\mu \theta_{i}} \!&=&\! \frac{\lambda}{2} \, \left( \, \partial_{\mu}\phi \star \partial_{\theta_{i}}\phi
+\partial_{\theta_{i}}\phi \star \partial_{\mu}\phi \, \right) \; ,
 \\
T_{\theta_{i}\theta_{j}} \!&=&\! \frac{\lambda^{2}}{2}
\left( \, \partial_{\theta_{i}}\phi \star \partial_{\theta_{j}}\phi +\partial_{\theta_{j}}\phi \star \partial_{\theta_{i}}\phi \, \right)
-\delta_{ij} \, e^{-\frac{\theta_{i}^{2}}{2\lambda^4}} \, {\cal L}_{\phi} \; . \nonumber
\end{eqnarray}
The energy-momentum tensor $T_{AB}$ is covariantly conserved in DFR space, {\it i.e.}, it satisfies the continuity equation with the covariant derivative operator: $\nabla_{A}\star T^{AB}=0$. The conserved components
of energy-momentum tensor are $T^{0B}=\left( \, T^{00} \, , \, T^{0i} \, , \, T^{0\theta_{i}} \, \right)$, where $T^{00}$
is the energy density, $T^{0i}$ is the spatial momentum density of the Euclidean space and the
component $T^{0\theta_{i}}$ can be understood as the spatial momentum of the scalar field in the extra $\theta$-dimension.
Using the result in Eq. (\ref{TAB}), the conserved momentum component $T^{0\theta_{i}}$ is given by
\begin{eqnarray}
T_{0 \theta_{i}} = \frac{\lambda}{2} \, \left( \, \dot{\phi} \star \partial_{\theta_{i}}\phi + \partial_{\theta_{i}}\phi \star \dot{\phi} \, \right) \; ,
\end{eqnarray}
which is a component associated with the DFR, and it goes to zero at the commutative limit $\lambda \rightarrow 0$.
Using the symmetry of the tensors in Eq. (\ref{EqEinstein}), we can verify that it has 25 non-linear differential equations
with corrections in the $\theta_{ij}$ parameter by the Moyal product between $G_{AB}(x,{\bm \theta})$ and $R$.
The field equation in Eq. (\ref{EqEinstein}) contains the NC Einstein's tensor in 4D dimensions if we set $A=\mu$ and $B=\nu$.
%

As an application of this formalism, let us obtain the wave equation for the spin-2 particle in DFR gravity structure. The action in Eq. (\ref{SGDFR}) and the field equation in Eq. (\ref{EqEinstein}) are both non-linear and they
also contain the Moyal product series that makes it difficult to deal with calculations. Hence,
we make the weak field approximation where the metric $G_{AB}(x,{\bm \theta})$ can be written as
\be
G_{AB}(x,{\bm \theta})\simeq \Xi_{AB}({\bm \theta})+\kappa \, h_{AB}(x,{\bm \theta})\,\,,
\ee

\ni where $\Xi_{AB}({\bm \theta})$ is the metric in Eq. (\ref{metricG}), and the symmetric field $h_{AB}$ has the components
\be
h_{AB}(x,{\bm \theta})=\{ \, h_{\mu\nu}(x,{\bm \theta}) \, , \, h_{\mu \theta_{i}}(x,{\bm \theta}) \, , \, h_{\theta_{i}\theta_{j}}(x,{\bm \theta}) \, \}\,\,.
\ee

\ni We are concerned with the linear terms in the field equation. It is well known, and it is a direct demonstration, that under an integral operation, like the one in Eq. (\ref{SGDFR}), considering only two functions and thanks to the anti-symmetric property of the $\theta^{\mu\nu}$ parameter, the Moyal product becomes the usual product of these two functions.  Namely, for more than two functions, this is not true for any order of $\theta^{\mu\nu}$, and the Moyal product must be calculated carefully with all its derivatives until the desired order in $\theta^{\mu\nu}$, since $\theta^{\mu\nu}$ is a Planck scale object. Thus, we obtain the Ricci tensor and the scalar one in the first order of $h_{AB}(x,{\bm \theta})$,
\begin{eqnarray}
R_{AB} \!\!&\simeq&\!\! \frac{\kappa}{2} \left( \partial_{A}\partial_{B}h - \partial_{A}\partial^{C}h_{BC} - \partial_{B}\partial^{C}h_{AC}
+\Box_{\theta} h_{AB} \right)  \; ,
\nonumber \\
R \!\!&\simeq&\!\! \kappa \left( \, \Box_{\theta} h - \partial_{A}\partial_{B}h^{AB} \, \right) \; ,
\end{eqnarray}
where $\Box_{\theta}:=\partial_{A}\partial^{A}=\Box+\lambda^2 \, \partial_{{\bm \theta}}^{\, 2}$
is the extended D'alembertian operator, and $h(x,{\bm \theta})=h_{A}^{\; \; A}(x,{\bm \theta})$.
As usual in effective gravity,
we choose the harmonic gauge, in which the $h_{AB}$ field satisfies the Donder gauge condition
$\partial_{A}h=2 \, \partial^{C}h_{AC}$. In this gauge condition, we obtain the wave equation
for $h_{AB}(x,{\bm \theta})$ for the case where we have no matter fields $(T_{AB}=0)$, namely,
\begin{equation}
\left( \, \Box+\lambda^2 \, \partial_{{\bm \theta}}^{\, 2} \, \right) h_{AB}(x,{\bm \theta})=0 \; .
\end{equation}
This is the wave equation concerning the components of $h_{AB}(x,{\bm \theta})$ with the propagation in the ${\bm \theta}$-space.
Note that the component $h_{\mu\nu}(x,{\bm \theta})$ that represents the spin-2 graviton propagates in the
extra dimension of the NC space. It would be interesting to investigate the contributions of this NC space to the
gravitational potential.  This is an ongoing research.


\section{Conclusions}
\label{sec4}
One of the main motivations to investigate NC space-time field models is that field  theories  with NC space and time coordinates give
an interesting opportunity to check  the possible breakdown of the standard concept of time and the well known
structure of QM at the Planck scale.
We proposed in this letter a gravitation structure for the Doplicher, Fredenhagen and Roberts NC framework.
The DFR model is a NC approach that includes the $\theta^{\mu\nu}$ parameter as a coordinate of the system, and consequently,
this formalism includes extra dimensions on the space-time. To keep a unitary QFT in DFR formalism, we have defined
$\theta^{0i}=0$.   Thus, we have a space-like NCY with three extra coordinates, {\it i.e.}, $(\theta_{12},\theta_{13},\theta_{23})$,
and the DFR space-time has seven dimensions: four from Minkowski space-time plus three spatial coordinates ${\bm \theta}$.
The product between fields defined in this space can be constructed via MW product, which is similar to the usual NC canonical formalism,
where $\theta^{\mu\nu}$ is a constant parameter.

We suggested a metric tensor for this extended space with the Minkowski components
together with Gaussian components associated with the weight function $W({\bm \theta})$ defined in the NC action.
The commutative limit is recovered using the Dirac delta identity in Eq. (\ref{DeltaID}). After that, we substituted the
Minkowski term by the metric $g_{\mu\nu}(x)$, defined on a smooth manifold that represents a NC space-time. Thereby,
we constructed a gravitation model with extra spatial dimensions in the DFR approach. The action and the corresponding field equations are
given by Eqs. (\ref{SGDFR}) and (\ref{EqEinstein}), respectively.

As an application of the formalism, we used the linear approximation in the NC Einstein's equation where the
wave equation with the propagation in the ${\bf \theta}$-space is obtained for the components of weak field $h_{AB}(x,{\bm \theta})$.
As a perspective, the analysis of the action in Eq. (\ref{SGDFR}) as an effective gravitation model in terms of the field $h_{AB}(x,{\bm \theta})$ is
an ongoing research that will be published elsewhere.

\section*{Acknowledgments}

\ni  The authors thank CNPq (Conselho Nacional de Desenvolvimento Cient\' ifico e Tecnol\'ogico), Brazilian scientific support federal agency, for partial financial support, Grants numbers 313467/2018-8 (M.J.N.) and 406894/2018-3 (E.M.C.A.).  M. J. N. also thanks the Department of Physics and Astronomy at the University of Alabama for the kind and warm hospitality.


\end{document}